\documentclass[sigconf]{acmart}
\usepackage{siunitx}
\usepackage{subcaption}
\usepackage{graphicx}
\usepackage{booktabs} % For formal tables

\newcommand{\A}{\mathcal{A}}

\newcommand{\D}{\mathcal{D}}

\begin{document}
\fancyhead{}

\copyrightyear{2018}
\acmYear{2018}
\setcopyright{iw3c2w3}
\acmConference[WWW '18 Companion]{The 2018 Web Conference
Companion}{April 23--27, 2018}{Lyon, France}
\acmBooktitle{WWW '18 Companion: The 2018 Web Conference Companion,
April 23--27, 2018, Lyon, France}
\acmPrice{}
\acmDOI{10.1145/3184558.3186949}
\acmISBN{978-1-4503-5640-4/18/04}

\title{User Fairness in Recommender Systems}
\author{Jurek Leonhardt}
\affiliation{%
  \institution{L3S Research Center}
  \streetaddress{Appelstraße 4}
  \city{Hannover, Germany} 
}
\email{leonhardt@l3s.de}

\author{Avishek Anand}
\affiliation{%
  \institution{L3S Research Center}
  \streetaddress{Appelstraße 4}
  \city{Hannover, Germany} 
}
\email{anand@l3s.de}

\author{Megha Khosla}
\affiliation{%
  \institution{L3S Research Center}
  \streetaddress{Appelstraße 4}
  \city{Hannover, Germany} 
}
\email{khosla@l3s.de}

\begin{abstract}
Recent works in recommendation systems have focused on diversity in recommendations as an important aspect of recommendation quality. In this work we argue that the post-processing algorithms aimed at only improving diversity among recommendations lead to discrimination among the users. We introduce the notion of \emph{user fairness} which has been overlooked in literature so far and propose measures to quantify it. 
Our experiments on two diversification algorithms show that an increase in aggregate diversity results in increased disparity among the users.

\end{abstract}

\maketitle

\section{Introduction}
Most recommender systems typically learn from past user interactions and preferences to recommend items (movies, products etc.) to users. The success of a recommender algorithm is generally evaluated with the accuracy of its recommendations, that is, how well the algorithm predicts whether a user will like an item or not -- its utility.
The aspect of user fairness arises when the task requires to consider the disparate impacts of recommendations on some user classes. On the one hand it might be unfair to ignore wishes of a certain class of users while trying to improve diversity in recommendations, on the other hand it is equally unfair to users if there is a lack of newness in the items which are recommended to them. A more dire situation can be seen in a job recommendation site, where a slightly under-confident person might always click on jobs with lower salary and is consequently always recommended jobs with lower salary distributions irrespective of his qualifications.

The utility-fairness conundrum is central to all fairness based measures. Specifically, making the recommendations fair will always result in a certain decrease in utility of the system.
From the fairness perspective we recognize two major sources of unfair distributions in recommender systems. The first, and more obvious, is the skewed distribution caused by the recommendations of items to users. This causes unfairness in the marketplace where certain items are recommended only very infrequently or not at all. The second and more subtle source of skew is caused by post-processing algorithms that address marketplace unfairness.
Much of the previous work relates to improving (1) \emph{individual diversity}, in which the focus lies on providing diverse recommendations to the users, and (2) \emph{aggregate diversity}, which focuses on improving item diversity across all users. Though individual and aggregate diversity can be interpreted as improving fairness for users and items respectively, they do not explore other aspects of fairness like differential treatment of two users or two items. For example, in order to improve aggregate diversity, an online store might recommend highly rated items to a set of users who are potential buyers, say the rich users, while new items (whose quality cannot be judged) are only recommended to poor users. On the one hand the recommender system might be unfair to the set of poor users, on the other hand it introduces item disparity by recommending new items only to users who might not actually buy them. Consequently, in designing fairness measures one needs to consider fairness criteria that should not unfairly discriminate against a certain set of users.

In this work we quantify the user unfairness or discrimination caused by the post-processing algorithms which have the original goal of improving diversity in recommendations. We perform experimental analysis on MovieLens and provide evidence that diversity improving algorithms can lead to discrimination among users.

\textbf{Related Work.} There have been recent, though limited, works on fairness aspects of recommender systems. In \cite{serbos2017fairness} the authors examine fairness issues in \emph{package-to-group} recommendations. Specifically, when they recommend a package to a group of people, they posit that this recommendation is fair, i.e.\ every group member is satisfied by a certain number of items in their package. Notions of novelty and diversity in recommender systems, as well as measures to quantify them and methods to improve them have been described by various authors~\cite{szlavik2011diversity,paudel2016updatable,antikacioglu2017post,adomavicius2014optimization,matt2013differences}.

Optimizing only for diversity can adversely affect accuracy, resulting in irrelevant recommendations. In \cite{zhou2010solving} the authors describe a hybrid approach that combines the ranking of an accurate algorithm with the ranking of a diverse algorithm.

\section{Measuring User Fairness}
 For the present work we restrict our attention to the movie recommendation task and show that post-processing algorithms that only optimize for diversity improvement among recommendations cause discrimination among users. We will need the following notations.

\textbf{Notations}: Let $I$ denote the set of $m$ items which need to be recommended to the user set $U$ of $n$ users such that each user is recommended $k$ items. We aim to understand the fairness aspects of the procedures followed for post processing recommendations. Let $G=(U \cup I, E)$ be the weighted bipartite graph representing $n$ users by the vertices in $U$ and $m$ items by the vertices in $I$. Let, for each edge $(u,i) \in E$, $w(u,i)$ represent the preference score (predicted) of user $u$ with respect to item $i$. For any $u \in U$, let $R(u)$ be the set of recommended items. Let $R^{top}(u)$ be the set of top-$k$ items for user $u$ with the highest preference scores. We assume that $R^{top}(u) \neq \emptyset$ for all $u \in U$.

Below we propose two measures for estimating user discrimination caused by the diversity improving algorithms. We first define \emph{user satisfaction} as a function of the relative gain achieved by the user due to the actual recommendation with respect to the optimal recommendation strategy (from the user perspective) where only the items with the top scores are recommended. Our first measure then computes the Gini coefficient for user satisfaction. Similarly, for our second measure we compute the user gain in terms of how many of the the recommended items match the top-$k$ items. We again compute the Gini coefficient of these user gains.

\textbf{Score Disparity}: First, we define user satisfaction for a user $u$ as the ratio of the sum of the preference scores for the items recommended to $u$ to the sum of the preference scores for the top-$k$ items, i.e.\ $\mathcal{A}(u) = {\sum_{j\in R(u)} w(u,j) \over {\sum_{j\in R^{top}(u)} w(u,j)}}$. Note that $0 \leq \A(u) \leq 1$ for all $u \in U$. Now, similar to how the Gini coefficient is used to measure disparity among populations, we define \emph{Score Disparity} as
$$\D_S={\sum_{u_1, u_2\in U} |\A(u_1) - \A(u_2)| \over 2n \sum_{u \in U}\A(u)}.$$

\textbf{Recommendation Disparity}: We first compute the similarity among the recommended items to users and their top-$k$ items with respect to preference scores as $sim(u)= {|R(u) \cap R^{top}(u)| \over k}$.

We then compute the user disparity which we refer to as \emph{Recommendation Disparity} based on the above computed similarity scores as
$$\D_R={\sum_{u_1, u_2\in U} |sim(u_1) - sim(u_2)| \over 2n \sum_{u \in U} sim(u)}.$$

\section{Experimental Results}
The MovieLens\footnote{available at \url{www.grouplens.org}} dataset that we use for our experiments contains \num{100000} ratings on \num{1700} movies from \num{1000} users. We use traditional collaborative filtering algorithms to obtain a list of predictions for each user. We then apply the standard ranking algorithm to obtain every user's top-$k$ predictions. Our experiments aim to illustrate the trade-off between recommendation diversity and user fairness. We run two post-processing algorithms (for recommendation diversity) on a set of predictions obtained from traditional CF algorithms ($k$-nearest neighbors and non-negative matrix factorization).

\begin{figure}
	\caption{Applying the \emph{Random} post-processing algorithm ($k = 5$) to predictions from CF algorithms. Each data point resembles a value of $\ell \in \{10, 50, 100, 500\}$. The black squares are special cases where $k = \ell$, i.e. no post-processing at all.}
	\label{fig_results_random}
	\begin{subfigure}{.49\linewidth}
  		\centering
  		\includegraphics[width=\linewidth]{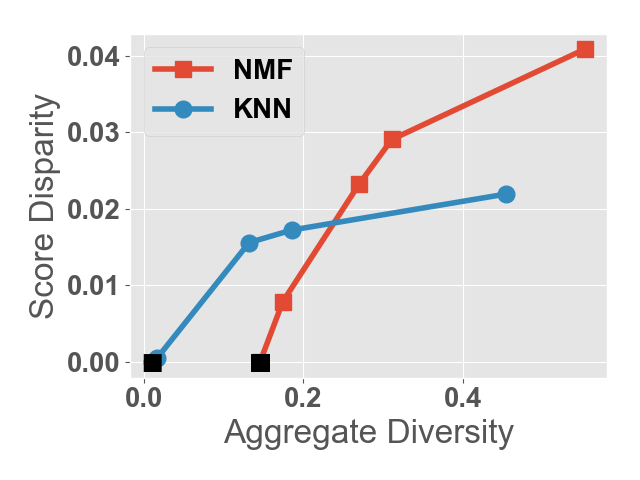}
	\end{subfigure}
    \begin{subfigure}{.49\linewidth}
  		\centering
  		\includegraphics[width=\linewidth]{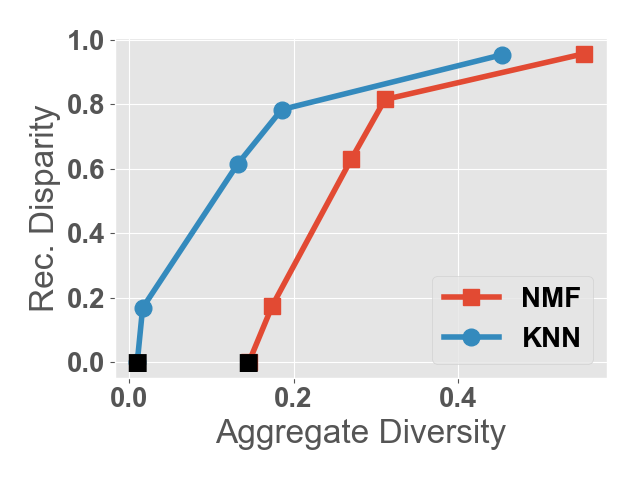}
	\end{subfigure}
	\\
	\caption{Applying the \emph{Greedy} post-processing algorithm ($k = 5$) to predictions from CF algorithms. Each data point resembles a value of $\theta \in \{10, 100, 200, 500, 1000\}$.}
    \label{fig_results_greedy}
	\begin{subfigure}{.49\linewidth}
  		\centering
  		\includegraphics[width=\linewidth]{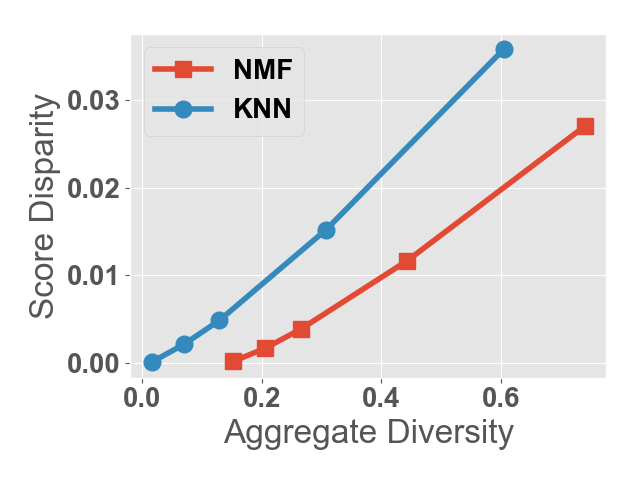}
	\end{subfigure}
    \begin{subfigure}{.49\linewidth}
  		\centering
  		\includegraphics[width=\linewidth]{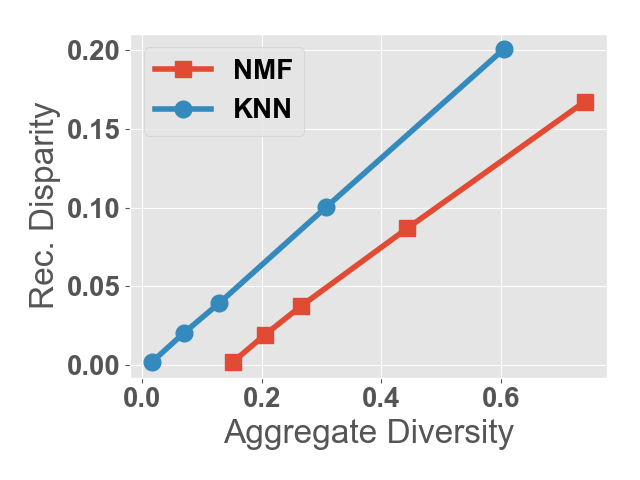}
	\end{subfigure}
\end{figure}

The first post-processing algorithm, referred to as \emph{Random}, takes a parameter $\ell$ and works by randomly sampling recommendations: To obtain a set of $k$ recommendations for a user, we simply employ the standard ranking algorithm to get the top-$\ell$ recommendations and then sample $k$ items uniformly from the result ($k \leq \ell$). This introduces some randomness to the final recommended items.
The second algorithm, also referred to as \emph{Greedy} (see Algorithm \num{1} from \cite{adomavicius2014optimization}), aims to increase the aggregate diversity among the recommendations by a value $\theta$ by replacing the top recommendations for a user with new recommendations (each having a preference score above a given threshold). We recall that aggregate diversity is defined as the fraction of total items which have been recommended at least once. The results of both algorithms are illustrated in Figures \ref{fig_results_random} and \ref{fig_results_greedy}. The plots clearly show an increase in the two defined user disparity measures when aggregate diversity increases. For example, applying the \emph{Greedy} algorithm to KNN-based recommendations (see Figure~\ref{fig_results_greedy}) causes an improvement in the aggregate diversity from $1.6\%$ to $60.5\%$ along with increases from $0.01\%$ to $3.6\%$ in Score Disparity and from $0.2\%$ to $20.1\%$ in Recommendation Disparity. This implies that there is indeed a trade-off between recommendation diversity and user fairness.

\section{Acknowledgements}
This work is partially funded by ALEXANDRIA (ERC 339233).

\begin{CCSXML}
<ccs2012>
<concept>
<concept_id>10002951.10003317.10003347.10003350</concept_id>
<concept_desc>Information systems~Recommender systems</concept_desc>
<concept_significance>500</concept_significance>
</concept>
</ccs2012>
\end{CCSXML}



\bibliographystyle{ACM-Reference-Format}
\bibliography{recsys} 

\end{document}